\documentclass[10pt]{article}
\usepackage{graphicx}

\def\Title#1{\begin{center} {\Large #1 } \end{center}}
\def\Author#1{\begin{center}{ \sc #1} \end{center}}
\def\Address#1{\begin{center}{ \it #1} \end{center}}

\newcommand\pubblock{\rightline{\begin{tabular}{l} Proceedings of the Fifth Annual LHCP\\ \pubnumber\\
         \pubdate  \end{tabular}}}

\newenvironment{Abstract}{\begin{quotation} \begin{center} 
             \large ABSTRACT \end{center}\bigskip 
      \begin{center}\begin{large}}{\end{large}\end{center} \end{quotation}}

\newenvironment{Presented}{\begin{quotation} \begin{center} 
             PRESENTED AT\end{center}\bigskip 
      \begin{center}\begin{large}}{\end{large}\end{center} \end{quotation}}

\def\Acknowledgements{\bigskip  \bigskip \begin{center} \begin{large}
             \bf ACKNOWLEDGEMENTS \end{large}\end{center}}

\def\beq{\begin{equation}}
\def\eeq#1{\label{#1}\end{equation}}
\def\eeqn{\end{equation}}

\def\beqa{\begin{eqnarray}}
\def\eeqa#1{\label{#1}\end{eqnarray}}
\def\eeqan{\end{eqnarray}}

\let\bar=\overbar

\def\Dslash{\not{\hbox{\kern-4pt $D$}}}
\def\dslash{\not{\hbox{\kern-2pt $\del$}}}

\def\msb{{\bar{\ssstyle M \kern -1pt S}}}

\usepackage{color}

\newcommand\pubnumber{ ATL-PHYS-PROC-2017-XXX }

\newcommand\pubdate{\today}

\def\affiliation{
Department of Physics and Astronomy and Center for Theoretical Physics \\
Seoul National University, Seoul, 08826, Korea}

\begin{document}

\large
\begin{titlepage}
\pubblock

\vfill
\Title{BSM LANDSCAPE}
\vfill

\Author{HYUNG DO KIM }
\Address{\affiliation}
\vfill
\begin{Abstract}

BSM landscape of motivated and unmotivated theories is overviewed with an emphasis on new developments guided by naturalness principle.
\footnote{"All science is either physics or stamp collecting." - E. Rutherford}


\end{Abstract}
\vfill

\begin{Presented}
The Fifth Annual Conference\\
 on Large Hadron Collider Physics \\
Shanghai Jiao Tong University, Shanghai, China\\ 
May 15-20, 2017
\end{Presented}
\vfill
\end{titlepage}
\def\thefootnote{\fnsymbol{footnote}}
\setcounter{footnote}{0}
%

\normalsize 


\section{Introduction}

What is going on at the weak scale and where is the particle physics going?

Particle physics was waiting for the discovery of new particles at the weak scale based on the guiding principle called naturalness to the hierarchy problem.
Higgs boson responsible for the electroweak symmetry breaking is the only spin $0$ scalar field in the the Standard Model and its lightness is the origin of the hierarchy problem.
Thus the Standard Model (SM) is incomplete by itself and needs a completion to address the problem. There have been a few popular candidates for the physics beyond the Standard Model (BSM).

The first and the most popular candidate is the weak scale supersymmetry.
The second is that Higgs boson is not a fundamental scalar but a composite particle.
It is connected to the idea of Higgs being a pseudo-Goldstone boson.
The common prediction of the models is that there are new particles at the weak scale in addition to the Higgs boson to keep the Higgs to be light.
Thus Large Hadron Collider(LHC) was expected to produce these new particles such that we could figure out the right explanation.

The long awaited Higgs particle has been discovered at the LHC in 2012 \cite{Aad:2012tfa} \cite{Chatrchyan:2012ufa}
and the Standard Model has been finally completed after 45 years.
However, on the contrary to the predictions made by most of beyond the Standard Model physics, there is no single evidence of new physics accompanied with the Higgs boson.

There is one important information we can learn from the Higgs boson discovery, the Higgs boson mass.
Most models beyond the Standard Model has a preferred Higgs boson mass unlike the Standard Model in which the mass itself is an unnatural free parameter.
Even in the Standard Model, the possible Higgs mass is limited from the stability ($> 100$ GeV) and the perturbativity ($< 1$ TeV) of the Higgs potential.

Dark matter and dark energy are two interesting topics as the SM cannot accommodate them. Dark matter is regarded as the topic of particle physics while dark energy is the most notorious problem that nobody provides a good explanation other than the anthropic one. Finally the charge quantization and running of gauge couplings indicate the unification of gauge group though the absence of proton decay makes the simplest unification model puzzling. Though there are many other motivations for BSM, let me focus on naturalness from now on as it was the main driving force of the particle physics for last four decades.

\begin{figure}[bht]
\centering
\includegraphics[height=2.5in]{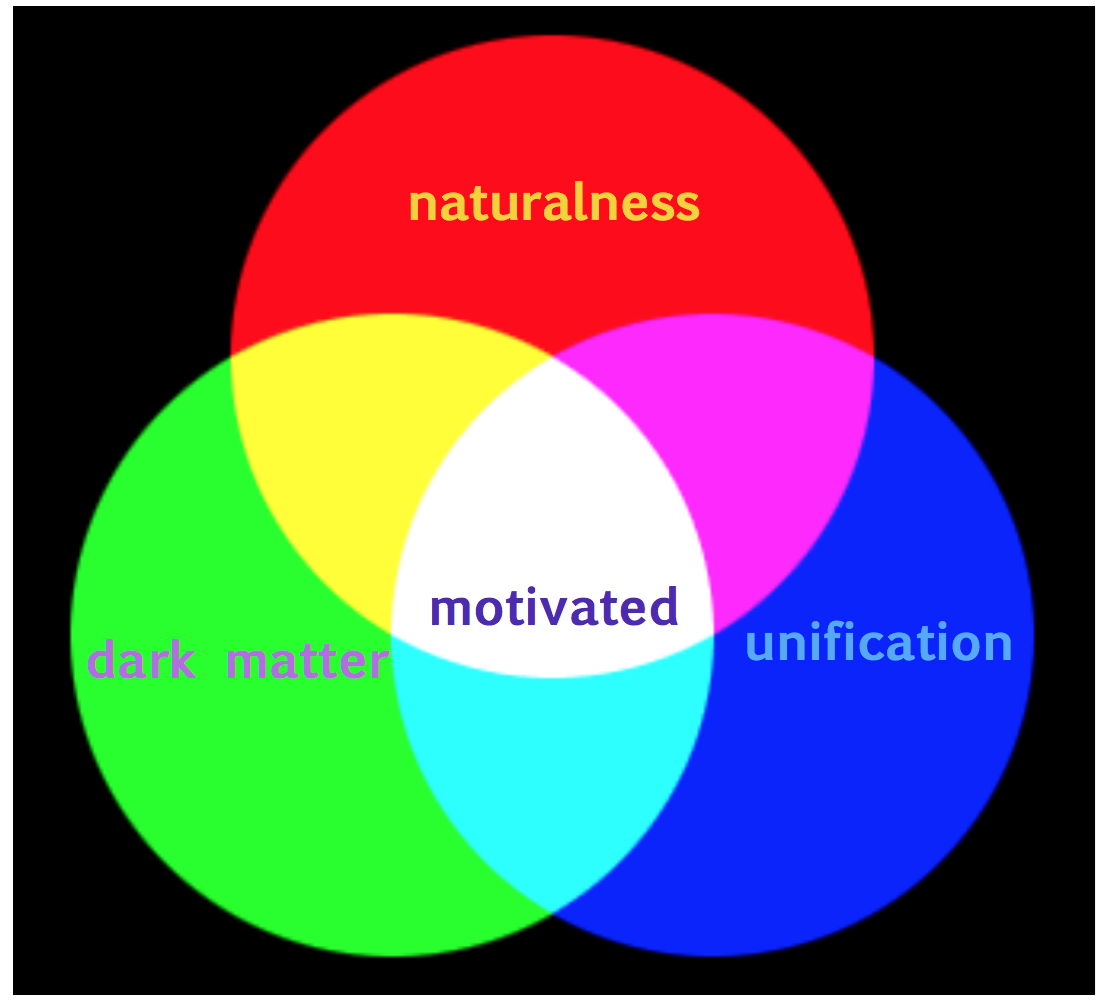}
\caption{Classically naturalness, dark matter and unification were considered as motivations for BSM.}
\label{fig:figure1}
\end{figure}


\section{Hierarchy problem and naturalness}

't Hooft naturalness criterion \cite{tHooft:1979rat} tells that the small parameter is natural if there is an enhanced symmetry at the point of vanishing parameter
as all the quantum corrections from the theory would be proportional to the symmetry breaking parameter and can remain small if the symmetry breaking is tiny.
This can be applied to the quarks and leptons (chiral symmetry) and W/Z bosons (electroweak gauge symmetry).
The photon remains massless as a result of the unbroken electromagnetic gauge symmetry.

The electron mass, the pion mass difference and the Kaon mass difference are three most successful examples of the naturalness principle.
In Quantum Electrodynamics(QED), naively the quantum correction to the electron mass is inversely proportional to the distance we can probe with the theory
and can be very large compared to the electron mass. However, this naive linear divergence of the quantum correction is exactly cancelled by the position contribution
and the correction is proportional to the electron mass itself and very mild logarithmic dependence is remained. Naturalness pre(post)dicted the positron.
This does not explain why the electron mass is smaller than W or Z boson mass by factor $10^{-6}$ but the stability of the physical quantity under the quantum correction is well explained.
In the Standard Model, the quarks and leptons are in the chiral representation and the mass term is generated only after the electroweak symmetry breaking of order $100$ GeV.
Thus there can be correction proportional to the largest symmetry breaking source, e.g., the top quark mass.
However, in the Standard Model, there is no leptoquark which can couple to electron and top quark at the same time \footnote{The author thanks R. Rattazzi for discussion on this.}
and as a result, the electron mass is proportional to the tree level electron mass itself and the radiative correction is guaranteed to be small if the electro mass is small.

The Higgs boson is the only (at least approximately) light fundamental particle with spin $0$ as far as we know \cite{Giudice:2008bi} \cite{Williams:2015gxa}.
The mass of the spin $0$ particle is unstable under radiative corrections.
There are several options to explain the lightness of spin $0$ particles.

\begin{enumerate}
\item Nambu-Goldstone boson

The mass of spin $0$ particle can not be protected unless it has a shift symmetry, in other words, the scalar particle is the Nambu-Goldstone boson.
The exact Nambu-Goldstone boson is massless and cannot have non-derivative couplings to preserve the shift symmetry.
The Higgs can be a pseudo-Nambu-Goldstone boson as long as the symmetry breaking terms are small.

\item Supersymmetry

Supersymmetry links the mass of the spin $0$ particle to spin $1/2$ particle whose mass is protected by chiral symmetry.
Supersymmetry should be broken since no supersymmetric partner has been observed.
However, the supersymmetric partner should be at around the weak scale to protect the Higgs mass and cannot be too heavy.

\item Compositeness

If the spin $0$ particle is composite, there can be no contribution beyond the compositeness scale and the hierarchy problem disappears.
The compositeness scale should be at around the weak scale for the lightness of the Higgs boson.

\item Landscape

If there are enough vacua, one of them might have the parameters which look unnatural otherwise.
Most of the attempts have been done for the cosmological constant problem which has more severe fine tuning problem
but it can be equally applied to understand the weak scale.

\item Cosmological solution

Though the Higgs mass can be very large in general, the Higgs boson mass can be settled down to a small value during the cosmological evolution.
Relaxion is one example in which the 'relaxion' field scans the Higgs mass parameter during its evolution and is settled down when the Higgs mass squared parameter becomes negative. N copies of the Standard Model with varying Higgs mass in each sector can also provide the solution if the 'reheaton' delivers most of its energy to the lightest Higgs sector (Nnaturalness).

\end{enumerate}

In the weakly interacting field theory, the exact dimension of the operator is close to the classical scaling dimension and we can classify the SM operators into the relevant, marginal and irrelevant operators using the classical scaling dimensions.
All of the SM gauge and Yukawa interactions are marginal operators and Higgs self coupling $|H|^4$ is also the marginal operator.
The Higgs boson mass term $|H|^2$ is the only exception and is the relevant operator with the parameter of mass dimension 2.
The hierarchy problem is the problem of quantum field theory with relevant operators.

\begin{eqnarray}
{\cal L_H} & = & {\cal L}_2 + {\cal L}_4  \nonumber \\
{\cal L}_2 & = & m^2 |H|^2 \nonumber \\
{\cal L}_4 & = & \lambda |H|^4 \nonumber
\end{eqnarray}

Before the Higgs discovery, the vacuum expectation value (VEV) of the Higgs boson was known and the ratio $\frac{m^2}{\lambda}$ was fixed.
The physical Higgs mass at the vacuum is
\begin{eqnarray}
m_h^2 = -2m^2 = 4\lambda \langle H \rangle^2. \nonumber
\end{eqnarray}

The quartic coupling $\lambda$ is a free parameter in the SM and can take any value.
Thus the physical Higgs mass is a free parameter in the SMl.

However, most of BSM predict the quartic coupling. Different models predict different ranges of the preferred Higgs mass.
On the contrary there is no theory predicting the quadratic term and the term is adjusted to provide the physical Higgs mass 
by the cancellation of the bare mass term and the calculable corrections if the quartic coupling is fixed.
\footnote{If both of them are calculable and predicted, it gives a wrong prediction on the Higgs VEV and is ruled out.}
Calculable corrections are much bigger than the needed physical Higgs mass and it is the fine tuning problem of the Higgs mass or electroweak symmetry breaking.

Let us write down all the couplings as dimensionless, $c_2 = \frac{m^2}{\Lambda_{\rm UV}^2}$ where $\Lambda_{\rm UV}$ is the UV cutoff of the theory.
Integrating out the high energy modes, $c_2$ is enhanced by a factor $\frac{\Lambda_{\rm UV}^2}{\Lambda_{\rm IR}^2}$ and the term becomes more and more important as we go to low energy. the operator of these kinds is the relevant operator.
If $H$ has no other interaction, the accidentally small $c_2$ would be enough to explain the smallness of $H$ mass.
\footnote{One possible way out is to make the SM Yukawa and gauge couplings to be relevant.}

In the Standard Model, $H$ has an order one Yukawa coupling to top quark and the renormalization group equation of $c_2$ is entangled with marginal couplings.
The resulting low energy $c_2$ becomes very sensitive to the change of high energy $c_2$.
The different scales mix and the separation of scales does not work  \cite{Williams:2015gxa}.

\subsection{Conventional approach I :  Weak scale supersymmetry}

In supersymmetry, the Higgs mass is tied up to the Higgsino mass $\mu$ and is protected against the quantum corrections once it is kept to be small.
$\mu$ is generated only after supersymmetry breaking and thus it can be linked to other soft supersymmetry breaking parameters.

Fermion mass can be protected by the chiral symmetry and light fermion mass is natural according to 't Hooft criterion.
Supersymmetry protects the scalar mass by relating its mass to the fermion mass.

If supersymmetric partners are at around the weak scale, the lightness of the Higgs boson is well understood.
In addition, in its minimal supersymmetric extension of the Standard Model (MSSM), the Higgs quartic couplings are linked to the gauge couplings by supersymmetry.
Only after the supersymmetric particles (top squark) are integrated out, the deviation is determined from the top quark loop and is logarithmically proportional to the mass separation of top squark and top quark in the leading order.
If the top squark mass is at around the top quark mass, the predicted upper bound on the Higgs mass is $m_h^2 \le M_Z^2$.
To increase the Higgs mass from $91$ GeV to $125$ GeV, the top squark should be at around $5 \sim 10$ TeV.
Then the heavy top squark give huge corrections to the quadratic terms in the Higgs potential
and this raised the question of why the weak scale is so much different from the top squark mass.

Before the Higgs discovery, most supersymmetric theories predicted the discovery of the Higgs boson at LEP, certainly below $115$ GeV.
The preferred range of the Higgs mass in the MSSM was at around $M_Z$, e.g., $100$ to $110$ GeV and even $115$ GeV was on the edge.
Currently observed Higgs mass $125$ GeV implies a fine tuning of order $10^{-3}$ to $10^{-4}$ in the MSSM with $5 \sim 10$ TeV top squark
and the result is not so much changed even if we consider large soft tri-linear $A_t$ term.
\footnote{When $A_t$ is large, the Higgs boson mass can be explained even with $1$ TeV top squark but this does not imply that the fine tuning is reduced by factor 25 or 100
as there are additional problem from $|A_t|^2$ and the net fine tuning is reduced only by factor 2 or 3 \cite{Kim:2012uy}.}
Furthermore, large soft tri-linear $A_t$ term is difficult to realise and needs a tachyonic boundary condition in UV theory \cite{Dermisek:2006ey}.

The prediction on the Higgs mass from the spectrum of the supersymmetric particles is sharp in the MSSM and can be relaxed if extensions including the singlet superfield(s) are considered. As there can be new quartic couplings solely coming from the interactions of the singlet with the Higgs doublets, the Higgs boson can be much heavier.
However, this enhancement is effective only when the VEVs of up type Higgs $H_u$ and down type Higgs $H_d$ are comparable. In the interesting parameter space which can increase the Higgs boson mass, the generic prediction is that the Higgs coupling to W/Z bosons, quarks and leptons are significantly modified compared to the SMl.
Adding another information that we not only measured the Higgs mass at $125$ GeV, the Higgs boson looks very much like the one in the SM, with a precision of order 10\%.
Thus, no significant deviation of the Higgs couplings compared to the SM lowers the chance that the extended models of supersymmetry including the Next to Minimal Supersymmetric Standard Model (NMSSM) would explain the Higgs mass with supersymmetric particles below 1 TeV.

Barring the extended supersymmetric models, the MSSM predicts $5$ to $10$ TeV top squarks and implies the fine tuning of order $10^{-3}$ to $10^{-4}$.
1 TeV top squark with $2 \sim 3$ TeV $A_t$ corresponds to $1.7 \sim 2.3$ TeV top squark without $A_t$ term as along as fine tuning is concerned and $10^{-3}$ fine tuning is unaviodable.
Natural supersymmetry was an attempt to understand serious direct search bounds on gluino and squarks while keeping the fine tuning as small as possible \cite{Asano:2010ut} \cite{Papucci:2011wy} .
With the discovery of the Higgs boson, natural supersymmetry setup does not help reduce the fine tuning as the constraint obtained to explain the observed Higgs mass in the MSSM is already too heavy and exceeds all the direct search bounds.

For supersymmetry believers, there are good and bad at the same time.
$125$ GeV Higgs mass needs at least $10^{-3}$ fine tuning in the MSSM and the scalar top quark is expected at a few TeV $\sim 10$ TeV.
On the other hand, the heavy scalar top quark predicted from the Higgs mass is very much consistent with no observation of supersymmetric particles at the LHC.

It is puzzling why there is $10^{-3}$ or $10^{-4}$ fine tuning if supersymmetry is the natural explanation for the weak scale.

\subsection{Conventional approach II : Compositeness}

Pions in QCD and Cooper pair in superconductor are two well known physical examples of light scalars with spin $0$.
They are not fundamental scalar fields and the ultraviolet (UV) theory is described in terms of fermions, i.e., the up/down quark and electron.

What if Higgs boson is low energy output composed of the fermions in the strongly interacting UV theory?
The question is legitimate and many alternative ideas other than supersymmetry are based on it.

The simplest try was to extend the gauge group to technicolor. Electroweak symmetry breaking is understood as a condensate of the techni-quark bilinear.
As it is possible only in a strongly interacting theory, it generically predicted the strong quartic coupling of the Higgs boson and heavy Higgs boson mass 
($\propto \lambda \langle H \rangle \sim$ TeV) and the notion of the scalar particle is inappropriate due to large decay width. 
As we disovered the light Higgs boson, this possibility has been falsified.

Still the chance to have the light scalar exists if it is combined with the idea of pseudo-Goldstone boson like pions in QCD \cite{Agashe:2004rs} \cite{Panico:2015jxa}.
However, the top Yukawa coupling is order one and top quark loop contributes sizeably to the Higgs mass.
In order to reduce the correction, the top quark partner $T^\prime$ is needed and its mass determines the fine tuning.
For the compositeness scale $f$, we expect techni-$\rho$ meson $\sim g_* f$ not very far from the weak scale.

Writing down the effective Lagrangian for the pseudo-Goldstone Higgs, we obtain

\begin{eqnarray}
{\cal L} = a \cos (\frac{h}{f}) + b \sin^2 (\frac{h}{f}). \nonumber
\end{eqnarray}

For $a \sim b$, the potential is minimised at around $\langle H \rangle \sim f$ and special tuning is needed to adjust $\langle H \rangle < f$.
This is the $v^2/f^2$ problem.
If nature was kind and the holographic composite Higgs boson was realised, then we would expect $\langle H \rangle \sim f$
and sizeable modifications of the Higgs couplings of order $\frac{v^2}{f^2}$.

The fine tuning problem becomes worse as the experimental constraints push $f \ge 700$ GeV and roughly 10\% fine tuning is inevitable.
The composite Higgs idea does not extend the UV cutoff in a straightforward way. Therefore, 10\% fine tuning in $v^2/f^2$ should not be treated in the same way as the fine tuning in supersymmetry in which the UV cutoff can be extended all the way up to the Grand Unified Theory (GUT) scale or the Planck scale.

\subsection{Radical approach}

All the models trying to understand the electroweak symmetry breaking have their own problems and there is no single preferred model.
It is related to the fact that most of the BSM predicts top partners at around the weak scale and we haven't found any new particle at the LHC.
What are the possibilities that we missed?

\subsubsection{Neutral naturalness}

The strong constraints obtained from the LHC is mostly on the colored particles as the LHC is the hadron machine.
If the quantum correction is cancelled by the color neutral particles, the constraints from the LHC would be very mild.
Extending $SU(2)_L$ to $SU(4)$ and introducing the mirror $SU(2)_L$ can realise the idea easily  \cite{Chacko:2005pe}.
Higgs boson comes as the pseudo-Goldstone boson of $SU(4)$ symmetry breaking.
The light Higgs would be generically predicted to be an half and half superposition of $SU(2)_L$ doublet and its mirror
which is drastically different from what we have observed at the LHC.
It is in parallel with the holographic composite Higgs and the main difference comes from color neutrality of the top partner.
In order to make the model to be consistent phenomenologically, the virtue of the original idea is gone
as the Higgs boson mass should appear as a cancellation of $SU(4)$ symmetry preserving (cutoff size) mass and the explicit $SU(4)$ breaking mass.
It predicts $f \sim v$ like composite Higgs (indeed it is a composite Higgs), but $\frac{v^2}{f^2}$ should be small to keep the Higgs couplings to be Standard Model like.
Fine tuning of order $\frac{v^2}{f^2}$ is unavoidable and also it is a theory with a lower cutoff unlike supersymmetry. Cosmology needs extra complication.
\footnote{The best summary of the approach was made by G. Giudice at CERN-CKC TH Institute in 2016. "It is brilliant and pathetic."}

\subsubsection{Relaxion}

There is no singled out explanation for the electroweak symmetry breaking as most natural parameter space is ruled out by experiments in all models.
The hierarchy problem itself originated from the relevant operator ${\cal L}_2$ and shares the common feature with the most notorious cosmological constant problem.
The cosmological constant is more relevant and has more serious fine tuning issue.
Widely accepted explanation is the landscape or the cosmological relaxion \cite{Weinberg:1988cp}.
It requires at least $10^{120}$ vacua with scanning cosmological constant
or considers the cosmological constant as the time varying parameter rather than the constant with suitable relaxation mechanism.
The cosmological constant problem might be infinitely more challenging than the particle physics problem
but still it would be an interesting try to apply the idea to the Higgs mass.

Relaxion linearly couples to $|H|^2$ and scans the Higgs mass. Initially large positive mass squared decreases monotonically as the relaxion rolls
and eventually becomes negative \cite{Graham:2015cka}. If Higgs mass squared changes sign, the quarks become massive
and the instanton potential barrier is generated.
In order for the mechanism to work, the decay constant $f$ of the relaxion should be much larger than the cutoff.
The large excursion of the fields would be possible in the clockwork setup \cite{Choi:2015fiu} \cite{Kaplan:2015fuy}.

\subsubsection{Nnaturalness}

\begin{figure}[htb]
\centering
\includegraphics[height=2.5in]{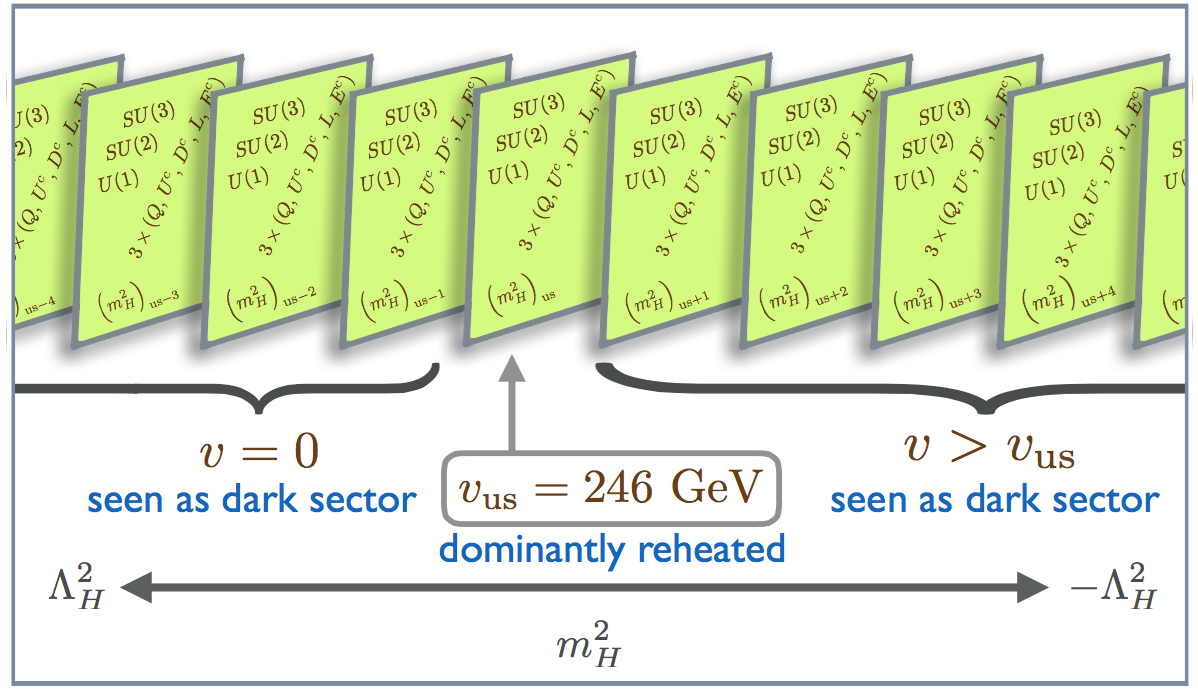}
\caption{Nnaturalness : N copies of the SM from \cite{Arkani-Hamed:2016rle}}
\label{fig:figure2}
\end{figure}

Landscape with sufficiently many vacua with scanning parameters can provide a comfort in the puzzling situation.
Nevertheless, there would be no observational evidence and we will never prove or falsify the explanation.
\footnote{"That which is not measurable is not science." - L. Kelvin \& E. Rutherford}

More concrete hybrid setup of naturalness and landscape is proposed in \cite{Arkani-Hamed:2016rle}.
Suppose there are $N$ copies of the Standard Model with scanning Higgs mass.
When $N \gg 1$, there are two interesting consequences.

\begin{itemize}
\item Lowering the Planck scale, $M_*^2 \simeq  \frac{M_{\rm Pl}^2}{N}$ : 

$N$ degrees of freedom modifies the scale at which the gravity couples strongly \cite{Dvali:2007hz}.

\item Lightest Higgs mass, $m_h^2  \simeq  \frac{\Lambda^2}{N}$ : 

With the simple assumption that Higgs mass is scanning from the cutoff $\Lambda^2$ to $-\Lambda^2$ with flat prior, the smallest negative mass squared would be $\sim 1/N$ smaller than the cutoff.
\end{itemize}

There are two represenative choices for $N$.

\begin{itemize}
\item $N=10^4$ MSSM with supersymmetry breaking at $10$ TeV: 

For $\Lambda = 10$ TeV, $m_h \sim 100$ GeV is well understood if $\mu$ parameter is scanning.
For $M_{\rm Pl} = 2 \times 10^18$ GeV, $M_{\rm GUT} = 2 \times 10^{16}$ GeV becomes the scale at which all three gauge couplings are unified and at the same time gravitational coupling becomes order one.

\item $N=10^{16}$ Standard Model: 

Now the scale at which the gravity couples strongly coincide with the UV cutoff, $M_* = \Lambda = 10^{10}$ GeV.

\end{itemize}

If all the sectors are reheated after inflation, we would end up with lots of dark radiation and dark matter inconsistent with the current cosmological observation.
If most of the energy density of the universe were carried by the field called 'reheaton' after inflation, the amount of dark radiation and dark matter can be calculated. 
If the reheaton couples to each sector universally and is light enough, then the relevant operator $ \phi |H_i|^2$ explains the preferred decay to lightest Higgs sector
where $\phi$ is the reheaton and $H_i$ is the Higgs boson in the i th sector (or  marginal operator $S L_i H_i$ if the rehaton $S$ is a fermion and $L_i$ is the lepton).

Interestingly, the solution to the particle physics problem has a prediction that no discovery at the LHC and interesting predictions in cosmology including the effective neutrino number, $\Delta N_{\rm eff}$, deviation of the matter power spectrum.

The novelty is that pseudo-Goldstone boson is beautifully realised.
The idea does not work very well for the Higgs boson since top Yukawa coupling is of order one (explicit breaking is too large) 
and a top quark partner at the weak scale is needed to be consistent.
Reheaton does not have such a constraint.
Then it is naturally light and can decay to fermion pairs or dibosons which gives a suppression inversely proportional to the mass of the Higgs boson.

It might provide a cosmological explanation for a light scalar without top partners.
Reheaton and N copies with relevant operator in between provide a reason why the small relevant operators are favored.
To test it as a principle, we have to push it further. The dimension of $H^\dagger H$ can be even lowered by strong gauge interactions which is more favored in this mechanism. 
Thus this principle cannot explain why weakly interacting Higgs is favored over the strongly interacting one.
\footnote{I'm indebted to M. Strassler on this point.}

Like other models, generic prediction of Nnaturalness is order one dark radiation, $\Delta N_{\rm eff}$ which is already ruled out.
Depending on the mass of the reheaton, there are parameter space predicting $\Delta N_{\rm eff} \le 0.5$.
If LHC doesn't find any new particle and $\Delta N_{\rm eff}$ is observed at CMB S4, Nnaturalness would survive as one of the plausible explanations.


\section{Motivated vs. unmotivated physics}

For more than 40 years, most of the beyond the Standard Model physics has been driven by the naturalness guideline
and apparently it doesn't look good at this moment judged from the null result of the LHC and possible other experiments.
Many attempts have been made with the spirit of removing strong prejudice and possibly wrong guidelines.

As an example, Coleman-Weinberg Higgs provides an electroweak symmetry breaking starting from $\lambda |H|^4$ Lagrangian without the negative mass squared term  \cite{Dermisek:2013pta}.
The quartic coupling change its sign in the renormalization group running (down) in the presence of new scalar and mixed quartic coupling.
The theory can be made to be perturbative all the way up to the Planck scale. Classically scale invariant theories are wrong by $10^{32} (=\frac{M_{\rm Pl}^2}{M_Z^2})$
unless the Coleman-Weinberg prescrption, $m^2=0$ in \cite{Coleman:1973jx} is understood.

The absence of new discovery other than the Higgs boson draw people's attention to the "Lamppost principle".
Though there is no good reason why the new particle should be discoverable or reachable at the LHC, it would be useful to think about all the possibilities as we know nothing about the nature like John Snow. This humble approach includes 'hidden valley', 'Higgs portal', 'dark photon', $Z^\prime$, vector-like quarks and leptons, and lepto-quarks.
No signal from direct dark matter detection experiments also make the option for the dark matter diverse as the most beloved weakly interacting massive particle (WIMP) becomes less and less attractive by the experimental constraints.

\section{Conclusions}

The hierarchy problem certainly exists for the electroweak scale and the naturalness has been an excellent guiding principle in the 20th century physics.
Experimental results are puzzling and there is no clear explanation on what is going on at the weak scale.
The guiding light should come from the end of the tunnel.
Without the guide from the experiments, the theory explorations will diversely diffuse and will fade away.
I wish oldies but goodies are realised in nature or something beyond our imagination would be discovered.

All of 'unmotivated physics' make a contrast with 'motivated physics' guided by naturalness.
More polished expression for 'unmotivated physics' is 'empirically motivated physics'.
'motivated physics' is 'philosophically motivated physics' to make it precise.
\footnote{H. Murayama provided the polished expression.}

Lamppost principle was useful in preparing what might come out from the LHC.
No new physics at the LHC tells us that nature is more subtle than we expected.
We need better ideas than fading away with unmotivated BSM physics.

'Leave no stone unturned' is the most popular guiding principle these days as nobody can guide the right direction with certainty.
Nevertheless most of the stones unturned are the ones which should have never been turned for obvious reasons.
\footnote{I share the opinion with R. Dermisek.}

\Acknowledgements
I am grateful to R. Dermisek, J. Ellis, G. Giudice, H. Murayama, R. Rattazzi, M. Strassler and A. Wulzer for fruitful discussions.
This work was supported by the NRF of Korea grant, No. 2017R1A2B2010749.

\end{document}